\newif\ifdevmode
\devmodefalse

\documentclass[sn-mathphys,twocolumn, iicol]{sn-jnl}

\jyear{2021}%

\theoremstyle{thmstyleone}%
\theoremstyle{thmstyletwo}%

\theoremstyle{thmstylethree}%

\newcommand{\tg}[1]{\textcolor{blue}{TG: {#1}}}
\usepackage{placeins}
\usepackage{subfig}
\usepackage{booktabs}

\begin{document}

\title[ ]{Identifying the root cause of cable network problems with machine learning}


\author*[1,2,4]{\fnm{Georg} \sur{Heiler}}\email{georg.heiler@tuwien.ac.at}

\author[3]{\fnm{Thassilo} \sur{Gadermaier}}

\author[4]{\fnm{Thomas} \sur{Haider}}

\author[1,2]{\fnm{Allan} \sur{Hanbury}}

\author[3]{\fnm{Peter} \sur{Filzmoser}}

\affil*[1]{\orgdiv{Institute of Information Systems Engineering}, \orgname{TU Wien}, \orgaddress{
\city{Vienna}, \postcode{1040},  \country{Austria}}}

\affil[2]{\orgname{Complexity Science Hub Vienna}, \orgaddress{\city{Vienna}, \postcode{1080},  \country{Austria}}}

\affil[3]{\orgdiv{Computational Statistics}, \orgname{Institute of Statistics and Mathematical Methods in Economics of the TU Wien}, \orgaddress{\street{Wiedner Hauptstrasse 8-10}, \city{Vienna}, \postcode{1040},  \country{Austria}}}

\affil[4]{\orgname{Magenta Telekom (T-Mobile Austria GmbH)}, \orgaddress{\street{Rennweg 97-99}, \city{Vienna}, \postcode{1030},  \country{Austria}}}


\abstract{
Good quality network connectivity is ever more important.
For hybrid fiber coaxial (HFC) networks, searching for upstream \emph{high noise} in the past was cumbersome and time-consuming.
Even with machine learning due to the heterogeneity of the network and its topological structure, the task remains challenging.
We present the automation of a simple business rule (largest change of a specific value) and compare its performance with state-of-the-art machine-learning methods and conclude that the precision@1 can be improved by 2.3 times.
As it is best when a fault does not occur in the first place, we secondly evaluate multiple approaches to forecast network faults, which would allow performing predictive maintenance on the network. 
}




\maketitle

\section{Introduction}\label{sec:introduction}

Hybrid fiber coaxial (HFC) networks deliver internet connectivity
directly to end customers.
Unfortunately, their reliability can be poor \cite{Bischof2018, Grover2013}.
The network contains separate channels for up (US) and downstream (DS) signals.
The US signal of the HFC network refers to data that is transferred from the customer up to the central root node.
In contrast, the DS part refers to the opposite direction of the signal, i.e., commonly used for downloads from the internet.
In particular, for a problem related to the US channels, a fault usually affects only a single or limited group of customers.
It relatively quickly spreads in the whole region of the network named fiber-node area.
Therefore resolving such a problem fast and without disrupting connectivity further is essential.
However, at the partnering internet service provider (ISP), the field technicians currently perform a binary search to identify the root cause of the problem by disconnecting certain amplifiers.
This means that not only is a considerable amount of time spent searching for the device, which is the root cause of the incident, the search process itself temporarily disrupts the service for other customers.

The cable industry suggests using proactive network measurements (PNM) to diagnose problems.
But the sheer volume of proactive alarms overwhelms the technicians as PNM data generically suggest areas of improvement and not the root cause of a specific incident.

Over time, implicit business knowledge has been built up to define a rule by the partnering ISP, but so far could not be executed automatically. 
We use it: Largest transmission power change before the incident -- as a baseline when comparing our results.
We demonstrate that by developing machine-learning enhanced models, precision can be improved over this baseline.
This allows to 2.3 times better (measured by precision@1) direct the technicians and faster resolve high noise faults in the network.
Such faults are sometimes referred to as common path distortion (CPD).

This problem is particularly interesting as normal behavior is different for each network region.
The topological structure of the network as defined in Section \ref{hfc-arch}should be included in the modeling approach.

In principle, it would be even better if a fault could be predicted before a field technician needs to be dispatched and customers observe degraded or unavailable service.
Therefore, we develop a prediction pipeline for network faults to showcase the potential of predictive fault detection.

Our research question is (I) to evaluate whether machine learning enhanced models can steer technicians better to a given root cause of a high noise incident and (II) whether a future incident indicated by an overly high codeword error ratio can be predicted in advance.
We contribute a label generation process and data pipeline to train machine learning models and can advance 2.3 times over the baseline when applying machine learning models to the problem.
Furthermore, we contribute a prediction approach using machine learning to showcase the possibility of genuinely predictive incident handling.


\section{State of the art}
For a given high noise incident we use machine learning models to steer technicians to the root cause of the incident.

The scientific literature focuses on issues in the DS path of the signal \cite{Zhu2020, Zhu2017},
	identification of anomalies \cite{Zhu2020},
	prediction of hotline calls from incident tickets and telemetry \cite{cablemon, heinzHFC}
	spectral analysis of the telemetry data for fault detection \cite{Rice, Zhang2010},
	collection of better quality data \cite{Thompson2020} directly from the cable modems,
	generic network data analysis with neural networks \cite{Ofdm2020}.
Tool vendors in the industry offer software solutions for individual and manual spectral-analysis-based failure analysis for specific devices.
However, too many warnings are created.
Additionally, technicians are not guided to the root cause of an incident as these systems generate too much data to obtain detailed information for the whole network in real-time.

In addition to the US data used in \cite{cablemon} we furthermore utilize the DS PNM data in our study as features for the various models.
The publications \cite{heinzHFC, cablemon} are trying to predict customer interactions on the hotline (based on generic faults), whereas we identify the actual root cause for any US high-noise-related incident automatically.
The authors of the tool CableMon \cite{cablemon} observe that they can predict approximately 80\% of trouble tickets that would lead to a call.
Eckert \cite{heinzHFC} observes a similar result when using autoencoders.

However, here for the high noise root cause detection, we are in a different setting:
Instead of only identifying an anomaly, we need to exactly pin-point the root cause of a given high-noise incident where often many cable modems start to act anomalously at almost the same time.

\section{Problem description}
In the following Section follows a description of the topological architecture of HFC networks as well as physical details of the problem.

\subsection{HFC architecture}
\label{hfc-arch}

\begin{figure}
\centering
\includegraphics[width=\linewidth]{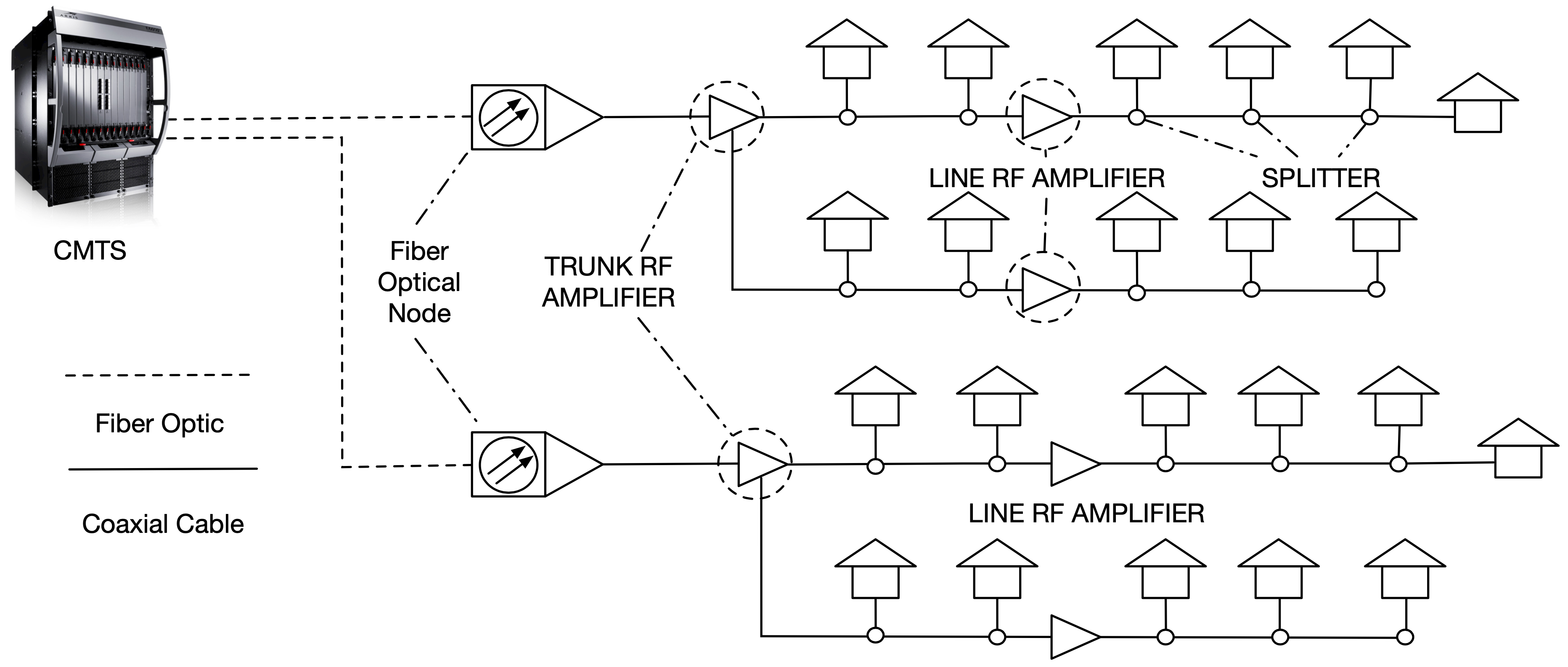}
\caption{Overview of the architecture of an HFC network from \cite{cablemon}. }
\label{fig:results:cablemon}
\end{figure}

The HFC network resembles a tree-like hierarchy.
An example is visualized in Figure~\ref{fig:results:cablemon}.
Often the network was built over a long period.
Usually, some operators were bought and merged in this process.
This contributes to further technical heterogeneity of the individual network segments (hubs).
Hubs represent the physical structure of the network region.
Commonly the devices in such a region were built together at the same time with the same technology and configuration.
Interestingly, some regions in the country are worse than others.
The root node of a hub named cable modem termination system (CMTS) contains several fiber-node areas which are connected using
optic-fiber.
Thus, these connections are highly reliable and in any case of failure, it is simple to identify the exact point of failure. 
The area of each fiber-node limits any signal interference.
A fiber-node - typically using many line- and distribution amplifiers and potentially splitters - connects the \emph{last mile} to the network.
The last amplifier before a final consumer, i.e., the house, is named the last line amplifier.
Based on coaxial copper cables, in particular, corrosion can badly influence the quality of the connections as parts of these networks are many decades old now \cite{cableHistory}.

PNM is recommended to improve fault resolution by the cable industry.
Monitoring tools deployed in the industry can generate many proactive alarms.
The sheer volume of proactive alarms can be overwhelming for the technicians.
Therefore, even though included in the Data Over Cable Service Interface Specification (\mbox{DOCSIS}) standard since 2005 \cite{clBestPractices}, dealing with PNM data remains a challenge as the recommendations for best practices and software deployed in the industry work with manually configured thresholds \cite{clBestPractices, Wolcott}. 
These are often used statically and tailored to use cases such as general proactive network maintenance.
Although the problems identified by PNM data indicate faulty network connections, they are not directly related to any specific customer disruption.
As a result, these identified problem notifications might deliver too many findings to be handled for a specific incident.
As there, the task is to identify the root cause quickly, given the limited human resources of the technicians.
Furthermore, the \emph{proactive} PNM alarms of a HFC plant monitoring system do not resemble any kind of predictions for maintenance, rather  only minor (=non-outage) faults, which could indicate the need for maintenance in the specific network elements if they frequently occur in a region of the network.
In other words: PNM data can be helpful to proactively gradually improve the quality of the overall network, but this data does not offer the clue for a specific incident.
In particular, PNM data does not outline which device is the root cause of any specific incident.
However, this information would be needed to guide technicians when the fault resolution process should be improved.
Accurate root cause indications have to be created with manual effort and this leads to the problem that this cannot be fulfilled with the available human resources of technicians.

\subsection{Fault characteristics}
\begin{figure}
\centering
\subfloat[Corroded F-connector]{\includegraphics[width=0.5\linewidth]{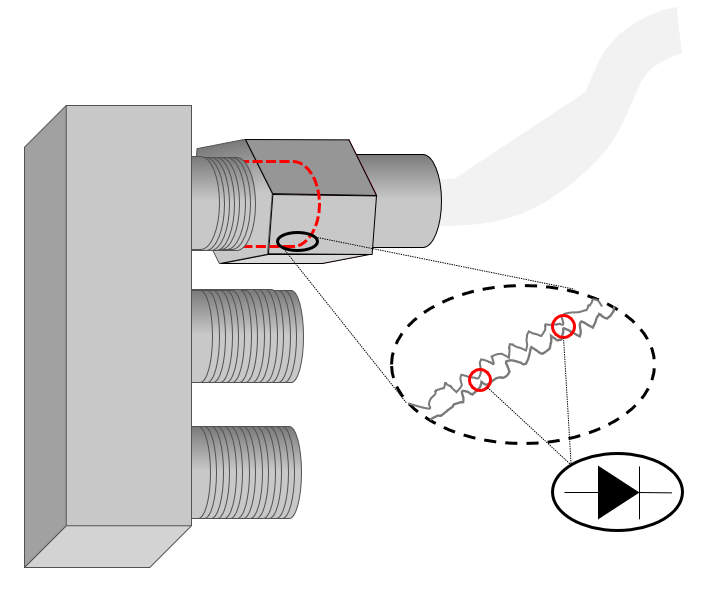}}\\
\subfloat[Loosely tied F-connector (partly due to corrosion)]{\includegraphics[width=0.5\linewidth]{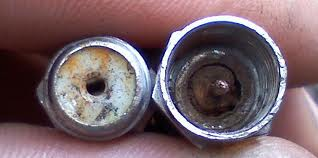}}
\caption{Upstream high noise is typically created by corroded contact surfaces or loosely tied connectors. Technicians of the ISP supplied these photographs of corrosion.}
\label{fig:brokenThings}
\end{figure}

High noise caused by CPD (Common Path Distortion) is an upstream distortion that is typically generated by corroded contact surfaces on a loosely tightened connector.
An example is shown in Figure~\ref{fig:brokenThings}.
For the specific high noise problem characteristics, it is essential to understand that US channels (i.e. frequency bands) are shared.
Therefore, a fault initially affecting only a single device on a specific frequency channel can quickly spread within the network region and in extreme cases destroy any connectivity in the whole fiber-node area.
Unlike downstream faults, where tracing these to a common specific point for technicians to fix, the upstream channel becomes more complex in case of problems as many cable modems can depict anomalous behavior in such a scenario at almost the same time.

In a normally functioning US channel, each cable modem sends the signal to a common point at the top of the cable network (CMTS) without any disruption.
The modems may not transmit on the same frequency at the same time.
The CMTS uses the \mbox{DOCSIS} protocol to control which modem is allowed to transmit at what time and frequency using the Time and Frequency Division Multiplexing (TaFDM) protocol. 
In case of disturbance on the US channel, any disturbance is transmitted onwards to the common point at the top of the network as coaxial cables are vulnerable to interference. 
Therefore, a single disturbance can negatively affect the \mbox{DOCSIS} signal for all modems in this fiber-node area or even make them unusable.
The downstream signal contains many different frequency bands.
These 
do mainly affect the US but also the downstream signal behind the corroded connector. 
\begin{figure}
\centering
	\includegraphics[width=\linewidth]{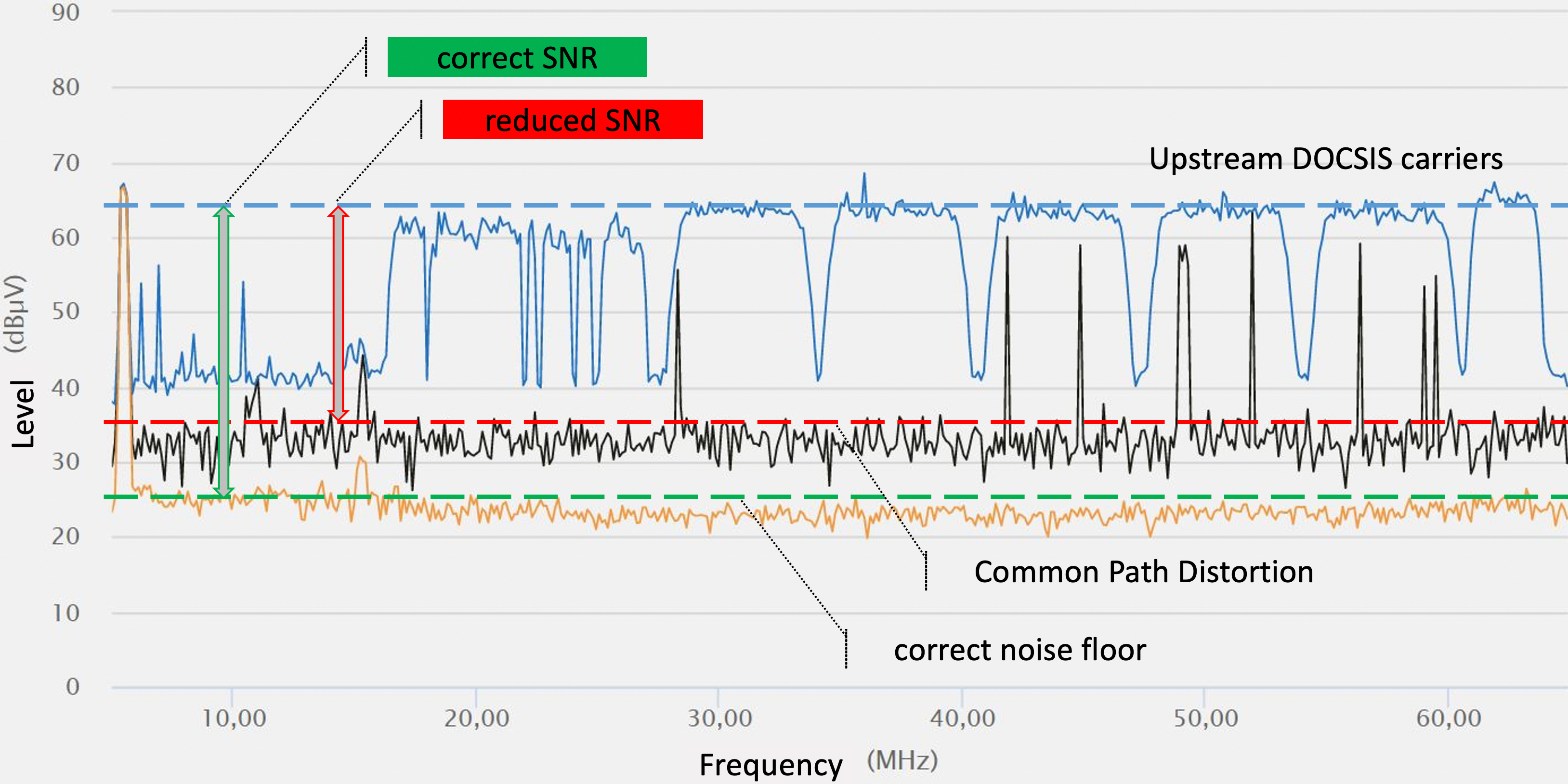}
\caption{
Only the upstream channel is visualized.
A noise floor is created by the vast amount of frequency bands participating in the incident. The x-axis is the frequency of the signal and the y-axis is the signal level for each frequency.
The yellow scenario (with the green marker) denotes a case with correct SNR, whereas the black scenario visualizes the noise floor (with the red marker) for reduced SNR.
The blue line represents the basic DOCSIS user data frequencies (carriers).
}
\label{fig:cpd_noise}
\end{figure}
In typical coaxial networks, this effect occurs at network points with a sufficiently high downstream signal where the US signal is relatively lower than the downstream.
Thus, the disturbance affects the upstream more than the downstream. 
The term \emph{High Noise} has become established as it forms a characteristic picture.
An example is found in Figure~\ref{fig:cpd_noise}.
It materializes as a noise floor in the spectral domain due to the huge amount of frequency bands participating in the fault.

Technicians are faced with the challenge of not knowing which point in the network the disturbance is coming from.
The current fault finding process is as follows:
A technician has to go through the network and identify where the fault is coming from on the path through the network to the root cause by conducting a binary search.
To make matters worse, the problem is often unstable and the technician cannot complete troubleshooting.
Only when the source of the problem is found, the process of fixing the fault can be initiated.
The main disadvantage is not only that technicians spend a lot of time troubleshooting, but that many customers are affected by the problem and that during the binary search procedure by the technicians to identify the root cause, additional customers might be affected.

In the following, we outline how the root-cause searching process can be improved by automating a simple rule-based classifier and utilizing machine learning enhanced methods.
Secondly, we present a fault prediction method to prevent faults from happening in the first place ideally.


\section{Methods}\label{sec11}
%
%

\label{sec:rca}
\subsection{Dataset description}


\emph{Telemetry data:} \label{p:telemetry} The telemetry information is collected on multiple levels: each cable modem reports data per each channel using simple network management protocol (SNMP) polling, but also the CMTS collects similar data.
However, the cable-modem-based data might not be available in network outages for particular modems.
The following are stored, in the raw form for each channel of each modem (MAC address), separately for both up- and downstream,
with hourly resolution including a timestamp:
signal to noise ratio
(SNR), a cable modem transmission power (Tx power), the received signal power (Rx power), codeword error ratio (CER) and corrected CER (CCER). 
For the downstream, additional micro reflections (m-reflection, impedance mismatch on the cable affecting the signal) are available.

\emph{Alarms:} The network operating center (NOC) stores alarming events for each device in Elasticsearch. 
For each device, the start and end of the alarm are noted.

\emph{Truckroll-tickets:} contain a free form text field for the notes of the technician, category of the incident, processing time and a free form text field for the amplifier causing the incident.

 We developed a parsing logic here to extract the 
 amplifier(s) which were identified as the root cause by the technician.
 Due to inconsistent naming of the amplifiers in different network regions and the process of parsing a free form text field, unfortunately, we are not able to utilize all tickets. 
 The tickets are filtered to contain high-noise-relevant tickets already only.

\emph{Topology:} Geospatial coordinates (location) for each 
amplifier as well as the 
path between the various amplifiers to the fiber-node.

\begin{figure}
\centering
	\includegraphics[width=\linewidth]{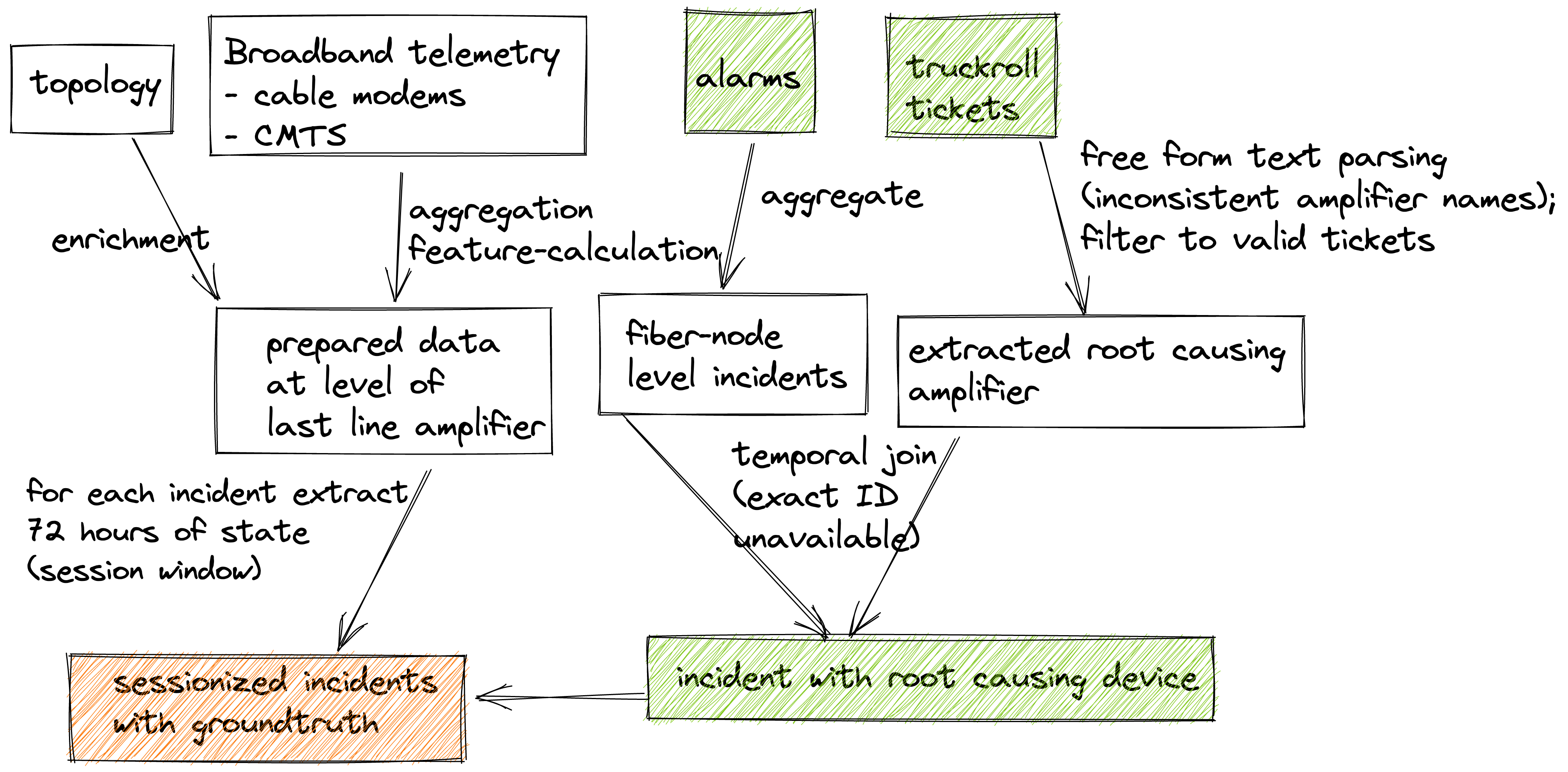}
\caption{Data pipeline overview. For each labeled incident, a dynamic state (session window) of 72 hours before each incident is obtained.} Using the available data, alarms and truck-roll tickets are correlated using a temporal join.
\label{fig:data_pipeline}
\end{figure}

The ground truth labels denote a root cause at a specific topology level.
We decided to only accept accurate root cause identifications as valid labels, which denote an individual amplifier (on the lowest level) as the root cause.
As the telemetry data is initially provided on the level of the fine-grained frequency bands where many belong to an individual amplifier, we decided to aggregate the data to the topological level of the last line amplifier.
Due to the sheer size of telemetry data for the whole country of the ISP we choose to use Apache Spark (version 3.1.2) \cite{Zaharia2012} to perform the aggregation.
During this aggregation process, the anonymity of the subscribers is ensured and we only ever receive anonymized data for our study.
Here, after linearly interpolating missing data for each device, we compute descriptive statistics (mean, std, min, max), change ratio (current/previous) and relative changes ((current-previous)/previous) for each feature.
Additionally, we consider a sliding window of 4 hours and calculate the change there as the difference between the largest and smallest value in each window instead of the difference between the current and previous observation.
This data aggregation process is depicted in Figure~\ref{fig:data_pipeline}.

We are evaluating a total of approximately five months of data (2021-02-25 -- 2021-07-25).
After the aggregation process, we consider 26069 last line amplifiers in the dataset, where some participate in multiple incidents.

An incident can become more severe (as more devices are affected).
We need to aggregate the individual device-level alarms to the whole fiber-node as high-noise-related incidents often affect many devices. We need to have the global beginning and end of the incident.
The global alarm time window is then used for a temporal join with the truck roll tickets as, unfortunately, no direct link between an incident, incident ticket, truck roll-ticket and the corresponding telemetry data was established before.
 Furthermore, only high-noise-related alarms are filtered for this specific use case.
 
\begin{figure}
\centering
	\includegraphics[width=\linewidth]{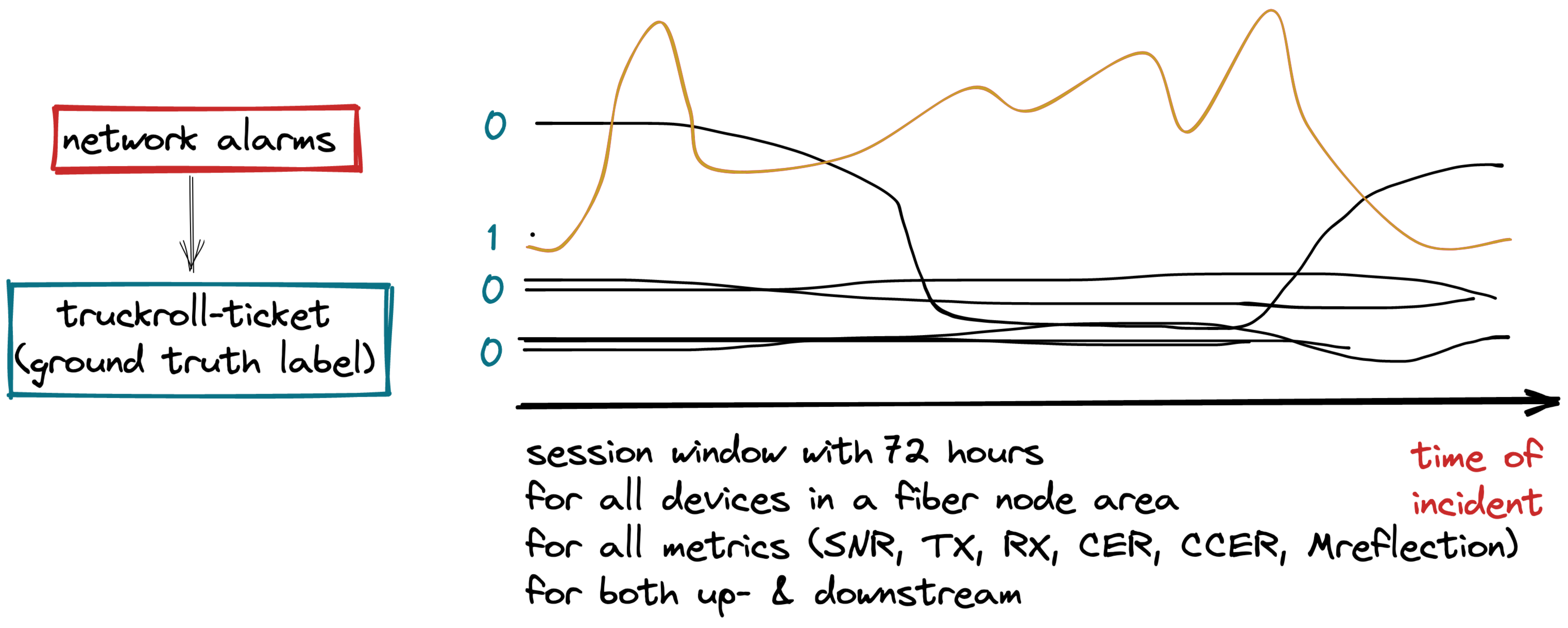}
\caption{Given network alarms (in red) and truck-roll tickets we perform a multidimensional sessionization on the telemetry data. 72 hours before an incident are kept as training data. The root cause label (defined by the truck-roll ticket as ground truth in blue) is used to identify the offending amplifier.
The y-axis contains the various amplifiers participating in the schematic incident session window.
The line of the amplifier causing the incident is highlighted in brown.
}
\label{fig:sessionization}
\end{figure}

With the alarms and parsed truck-roll tickets we can obtain ground truth labels for each incident.
For each incident, we obtain a session window where one or more last line amplifier is marked with the label denoting a root cause for this particular incident.
Figure~\ref{fig:sessionization} depicts an example case of the classical Tx spikes before a high noise incident that matches the positive class label.

As we need to identify the root cause for a specific incident, we can only keep incidents where a label is available in our dataset. 
796 root cause amplifiers remain labeled from the ground truth data from 7 network regions for 457 unique fiber-node areas and 672 truckroll tickets.
This means that for some tickets $\ge 1$ offending (= root-causing amplifier) are suggested in the ground truth data.
In total, we obtain 796 positively labeled amplifiers out of a total of 26069 for an amplifier identified as the root cause of a high noise incident\label{p:ticket_numbers}.
The remaining data are kept as negative examples.
This makes the dataset highly unbalanced with regard to the target labels.

\subsection{Data Preprocessing}
The network as described in Section~\ref{hfc-arch} contains two crucial levels in the network's topology: hubs and fiber-nodes.

For most statistical models, numeric distances between the features are important.
As the quality, if the network is different in each region differs, we must normalize the data in a way that we can learn from and compare all incidents taking both the physical effects and error boundaries mentioned above into account, as a feature that is considered high or anomalous in one region might be completely normal for another one.
We propose to double normalize the data: 
As discussed in Section~\ref{hfc-arch}, the devices in a hub have similar physical properties, which can be handled using simple standardization (0 mean, unit variance).
When introducing the HFC topology we already explained that any error is limited to the extent of each fiber-node area, Section~\ref{hfc-arch}.
To make amplifiers comparable across fiber nodes, we need to standardize again, now taking time into account and do so for each of the 72 hours of session window for each feature, but standardize only within the amplifiers related to this particular incident.


\subsection{Models}
Starting with a simple business rule as our baseline we compare various state-of-the-art (SOTA) ML/AI-enhanced approaches.

\textbf{Baseline: business rule}
Decades of knowledge of the technicians define a very simple business rule as follows:
Shortly before the incident, the largest upstream Tx change identifies the root causing amplifier.

This rule has a significant advantage: 
It is dynamically adapting to the specific situation of each incident due to choosing the largest change.
Given two very different network regions with different physical properties or quality, the largest change is still a fairly reliable indicator for the amplifier causing the incident.
Furthermore, this simple rule is well understood by the technicians.
In case fine-tuning is required, they can easily adjust the cutoff parameters for themselves and instead of analyzing the top-1 (largest) change they could consider the top-n.

As we will see later, when evaluating statistical machine-learning models, such dynamics which is specific for each incident needs to be explicitly considered there as well during evaluation\label{p:simple-dynamic}.

\textbf{Subgroup discovery}
Using explainable models can increase the trust of the non-tech business stakeholders as they can easily understand the inner workings.
Singh et. al. \cite{imodels2021} provide a package with implemented models that might be able to replace black-box models with simpler ones while improving efficiency and interpretability without sacrificing accuracy. 

\textbf{ML models}
\emph{Logistic regression:} A simple statistical baseline using a standard logistic regression procedure, it is implemented in scikit-learn \cite{scikit-learn}.
\emph{Lightgbm}
\cite{Ke2017} is one example for gradient boosted tree models which generally deliver good model fitting performance and is fast to train. Unlike neural networks, it does not require extensive fine-tuning.

We compare various neural network-based approaches as well. These are based on 
tsai \cite{tsai} as an implementation of various state-of-the-art time-series oriented architectures based on fast.ai \cite{Howard2020}.
We use the models of tsai for our dataset and in particular, adapt the data loaders for the sessionization and normalization as outlined above.
For any of the neural network models, we use the learning rate finder\footnote{\url{https://sgugger.github.io/how-do-you-find-a-good-learning-rate.html}}
provided by the fast.ai library to balance the speed of training and accuracy of the models whilst still improving the performance of the models as there is a smaller chance being stuck in local optima.
\emph{LSTM}
long-short-term-memory is a traditional neural network architecture for time series handling \cite{lstm}
\emph{InceptionTime}, is a recent SOTA architecture for time series \cite{inceptionTime}.
\emph{TST} BERT \cite{bertgoogle} and transformers revolutionized the field of sequence-based neural networks.
Only recently first adaptations of these models for temporal tasks have been developed. 
The time series transformer (TST) \cite{tst} is one such example.
It is based on \cite{Furfaritony2002, He2021} the domain of information retrieval.

Both text- and image-based domains were revolutionized when pre-trained models could be used.
This drastically decreased the required compute resources and datasets.
For the time-series domain, the classical pre-trained models cannot be used as they stem from a completely different domain.
Instead, we follow a self-supervised pre-training\footnote{\url{https://github.com/timeseriesAI/tsai/blob/main/tutorial_nbs/08_Self_Supervised_TSBERT.ipynb} \cite{tsai}} approach by first training a BERT based model in unsupervised mode to create network embeddings for our LSTM core; secondly, we use this pre-trained model in three scenarios:
\emph{LSTM self supervised (fine-tuning)}, \emph{LSTM self supervised (training)}, and \emph{LSTM self supervised (train) + data augmentation} training with the CutMix \cite{cutmix} data augmentation strategy.

The hyperparameters of the models were optimized using Optuna \cite{optuna} on a GPU-equipped server.

\subsection{(Ranked) evaluation of results}
When evaluating the models we do \emph{not only} perform a classical binary classification evaluation, where for one particular observation a probability is emitted.
Rather, we classify a single incident session globally by obtaining the predictions of the model if any amplifier is a root cause for the incident and then rank these predictions.
The ranked evaluation takes place in two stages: Firstly, the binary classification is performed by the various models.
Secondly, the output probabilities are ranked and a top-k evaluation is performed.
This is a deliberate decision as it enhances each of the models with the dynamics of the particular incident and network region as mentioned in Section \ref{p:simple-dynamic} we could not account for otherwise.
Empirically this proves to work well for all models, as we are in a ranking task ,where the most probable root cause for each incident needs to be identified, when analyzing the precision@k, see Table~\ref{tbl:results:rca}.




\section{Results}\label{sec2}


The \emph{discovered subgroups} can be used to create rules which are  easily understandable.
Interestingly, these statistically discovered rules align well with the business practice of the field technicians.
Showcasing the technicians that we can use these to derive their business rule increased trust in our other modeling activities.

\begin{figure}
\centering
\subfloat[Precision]{\includegraphics[width=0.9\linewidth]{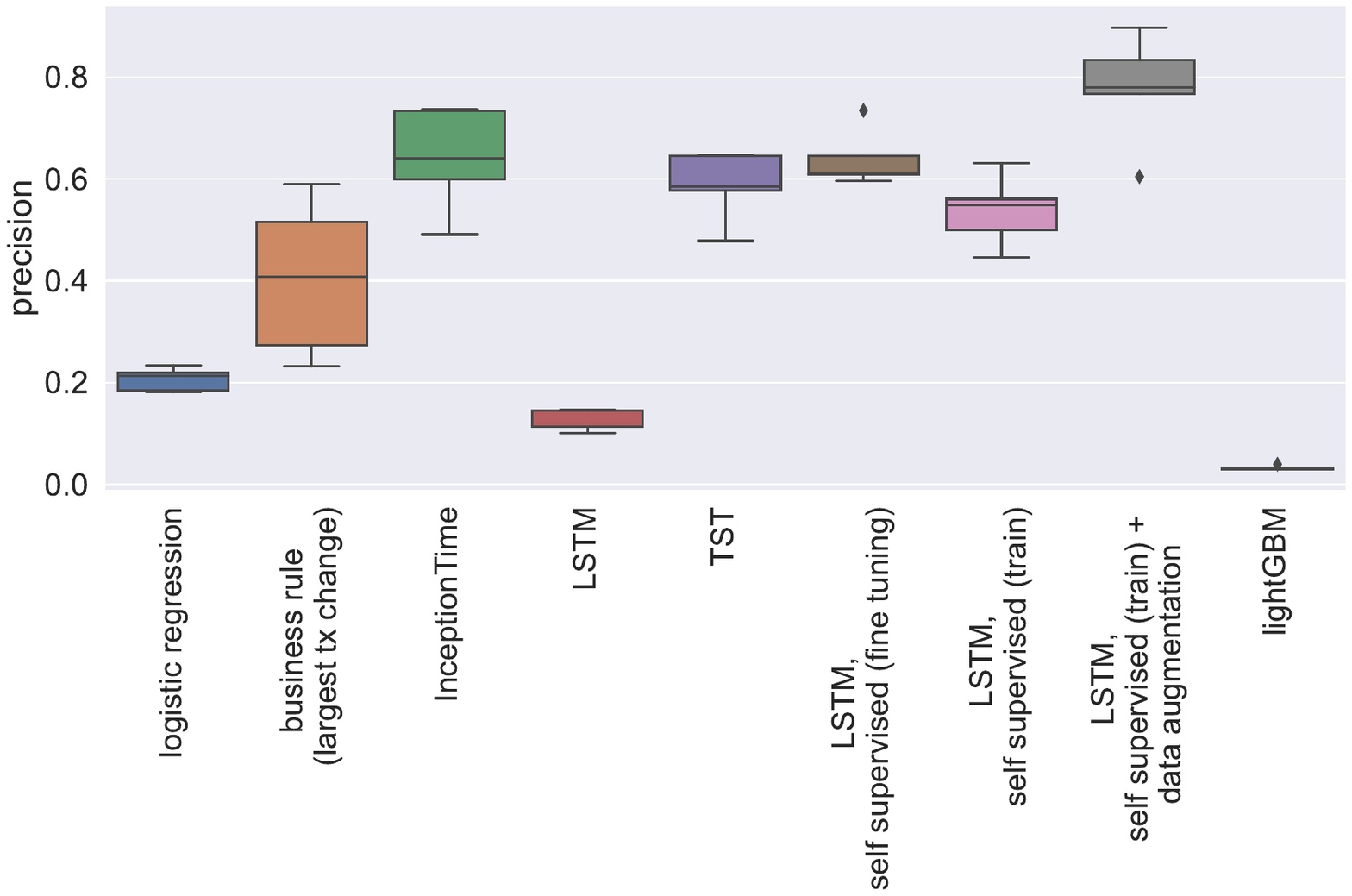}}\\
\subfloat[Recall]{\includegraphics[width=0.9\linewidth]{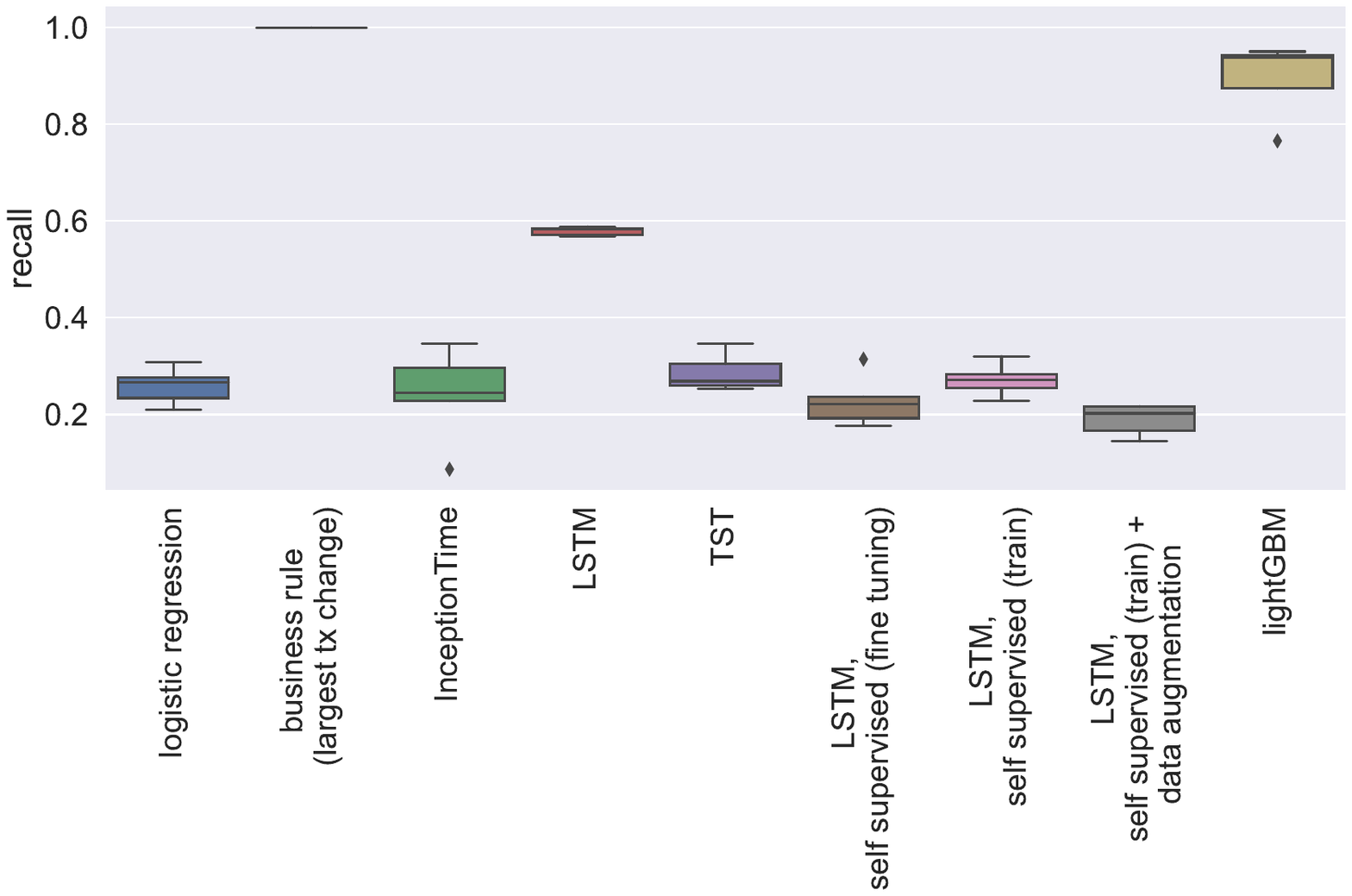}}
\caption{Precision and Recall for the raw model outputs of the first binary classification stage for each cross-validation fold.}
\label{fig:model_eval_precision_and_recall}
\end{figure}

The results of the first raw (binary classification) evaluation are depicted in Figure~\ref{fig:model_eval_precision_and_recall}.
The logistic regression is worse with regards to both precision and recall compared to the business rule.
Most of the other models (except LightGBM and LSTM) result in higher precision.
With regards to recall, none of the other models is better than the business rule.
However, the business rule baseline can achieve the high recall only with very limited precision.
In any real-world scenario when deployed at an ISP the human resources of the technicians are limited to evaluate false alarms, therefore a high precision is more important than recall as the technicians otherwise might lose trust in the technical solution.

\begin{figure}
\centering
\includegraphics[width=\linewidth]{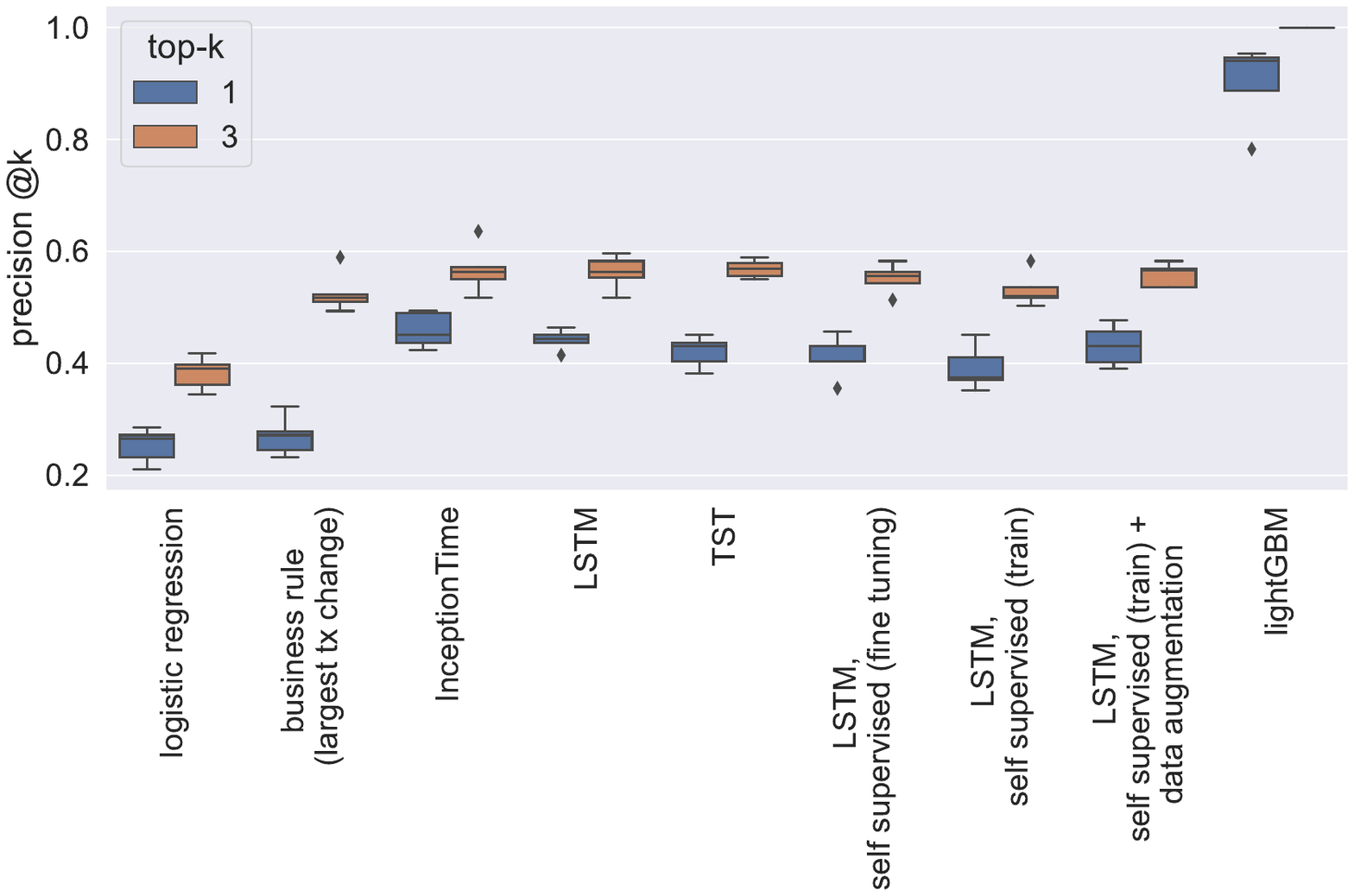}
\caption{Comparison of models using precision@k (ranked). Any of the more complex models deliver better results than the simple business rule for both top-1 and top-3 evaluation. LightGBM in particular performs best with a wide margin. Only a naive logistic regression is worse than the business rule.}
\label{fig:results:precision:new_business_ranked}
\end{figure}

Furthermore, when reframing the task into a ranking task where the most probable root cause is identified, the superiority of the ML enhanced models clearly becomes visible.
The models are evaluated for a top-1 and any within the top-3 match.
However, for sake of brevity and increased precision, we only discuss the top-1 match when comparing the results, as this is the variant that would most likely be used by an ISP to minimize the workload overhead of the technicians induced by faulty recommendations. 
Figure~\ref{fig:results:precision:new_business_ranked} depicts both cases for completeness.
Detailed results for precision, recall and the precision@rank-k are listed in 
Table~\ref{tbl:results:rca}.
\label{si:results:rca}
\begin{table*}
    \centering
    \resizebox{\textwidth}{!}{%
  \begin{tabular}{llrrrrrrrrrr}
\toprule
  &                     & \multicolumn{2}{l}{precision step 1} & \multicolumn{2}{l}{recall step 1} & \multicolumn{2}{l}{precision @k} & \multicolumn{2}{l}{false positives@k} & \multicolumn{2}{l}{true positives@k} \\
  &                     &             mean &    std &          mean &    std &         mean &    std &              mean &     std &             mean &     std \\
top-k & model &                  &        &               &        &              &        &                   &         &                  &         \\
\midrule
1 & lightGBM &            0.032 &  0.004 &         0.894 &  0.078 &        0.902 &  0.072 &              14.8 &  10.918 &            136.4 &  10.502 \\
  & InceptionTime &            0.640 &  0.103 &         0.241 &  0.098 &        0.459 &  0.031 &              81.8 &   4.604 &             69.4 &   4.879 \\
  & LSTM &            0.130 &  0.022 &         0.579 &  0.009 &        0.442 &  0.018 &              84.4 &   2.966 &             66.8 &   2.588 \\
  & LSTM, self supervised (train) + data augmentation &            0.777 &  0.109 &         0.189 &  0.032 &        0.431 &  0.036 &              86.0 &   5.612 &             65.2 &   5.404 \\
  & TST &            0.586 &  0.068 &         0.286 &  0.039 &        0.421 &  0.028 &              87.6 &   4.393 &             63.6 &   4.037 \\
  & LSTM, self supervised (fine tuning) &            0.639 &  0.056 &         0.229 &  0.053 &        0.415 &  0.038 &              88.4 &   6.066 &             62.8 &   5.675 \\
  & LSTM, self supervised (train) &            0.537 &  0.069 &         0.272 &  0.034 &        0.392 &  0.039 &              92.0 &   6.000 &             59.2 &   5.891 \\
  & business rule (largest tx change) &            0.270 &  0.035 &         1.000 &  0.000 &        0.270 &  0.035 &             110.4 &   5.030 &             40.8 &   5.404 \\
  & logistic regression &            0.206 &  0.023 &         0.259 &  0.038 &        0.253 &  0.031 &             113.0 &   4.899 &             38.2 &   4.550 \\
3 & lightGBM &            0.032 &  0.004 &         0.894 &  0.078 &        1.000 &  0.000 &               0.0 &   0.000 &            151.2 &   0.447 \\
  & TST &            0.586 &  0.068 &         0.286 &  0.039 &        0.569 &  0.016 &              65.2 &   2.387 &             86.0 &   2.550 \\
  & InceptionTime &            0.640 &  0.103 &         0.241 &  0.098 &        0.567 &  0.044 &              65.4 &   6.580 &             85.8 &   6.611 \\
  & LSTM &            0.130 &  0.022 &         0.579 &  0.009 &        0.562 &  0.031 &              66.2 &   4.658 &             85.0 &   4.583 \\
  & LSTM, self supervised (train) + data augmentation &            0.777 &  0.109 &         0.189 &  0.032 &        0.558 &  0.021 &              66.8 &   3.114 &             84.4 &   3.209 \\
  & LSTM, self supervised (fine tuning) &            0.639 &  0.056 &         0.229 &  0.053 &        0.552 &  0.026 &              67.8 &   4.087 &             83.4 &   3.715 \\
  & LSTM, self supervised (train) &            0.537 &  0.069 &         0.272 &  0.034 &        0.532 &  0.031 &              70.8 &   4.712 &             80.4 &   4.615 \\
  & business rule (largest tx change) &            0.526 &  0.037 &         1.000 &  0.000 &        0.526 &  0.037 &              71.6 &   5.683 &             79.6 &   5.459 \\
  & logistic regression &            0.206 &  0.023 &         0.259 &  0.038 &        0.382 &  0.029 &              93.4 &   4.506 &             57.8 &   4.324 \\
\bottomrule
\end{tabular}

  } 
    \caption{Summary statistics (mean, std) for the results of the various root cause analysis models. Notice: The counts are aggregated high noise incidents. Each incident contains a varying but high number of underlying amplifiers.
     \label{tbl:results:rca}}
\end{table*}
The simple \emph{business rule} (largest Tx change before the incident) results in a top-1 precision (on average for the cross-validation folds) of 0.27. 
Any of the other more complex models deliver better results: 
In particular the \emph{tree-based} model LightGBM results in a top-1 precision of 0.902.
The various neural-network-based approaches differ in their precision only marginally  (0.392 -- 0.45) and are additionally worse and more complex to interpret and computationally expensive to train than LightGBM.
Interestingly, LightGBM outperforms all the neural network-based approaches in our comparison.
Most likely the reason for this is that the amount and quality of the training data is limited so far:
\begin{itemize}
	\item \emph{quantity}: we were only able to obtain ground truth labels for a limited area in the network see Section ~\ref{p:ticket_numbers} due to free form text field parsing
	\item \emph{quality}: due to a missing id field in the various data sources connecting the alarm and field-force ticket to an incident we need to perform a temporal correlation.
\end{itemize}
LightGBM has the advantage that the model training procedure is swift and allows for more experimentation with regards to hyperparamters.
Especially in an industrial context where often AI is only an enhancing part of the overall process the optimization of the hyperparameters can be performed very fast.

%

\section{Discussion}\label{sec12}
%
%
%
%
%
%

The machine learning aspects are only part of a bigger use case.
Hence, it is important to understand the requirements of the ISP well in case it should be deployed in a scalable real-time setting with integration into an existing processing landscape.
Indeed, the simple automation of the presented business rule will have the advantage of being most easy to get started with and transparent to the HFC technicians maintaining and operating the network.
However, as we have shown any of the other more complex models outperforms this simplistic business rule by a wide margin: The best one, LightGBM, is more than 2.3 times better than the baseline.

To benefit most from these ideas, the ISP should further consider creating a heat-map of the identified root cause devices over a more extended period of time.
Thus if repeatedly problems are identified in an area, technicians can be dispatched there, perform maintenance and improve the overall quality of the network (not related to any specific incident).
This can be especially useful in case of hard-to-reproduce (flaky) problems. 

The suggested approach can be improved by making more and better quality data available:
\begin{itemize}
	\item generating more training data: Creating a structured ticket reporting instead of free form text field parsing.
	\item collecting better quality labels: Currently, a temporal join is required to link the telemetry data with the alarms and incident tickets. Instead one stringent equi-join (id-based) identification of incident and label would further enhance the quality of the labels available to the ML pipelines
\end{itemize}

In the long run, upgrading the infrastructure as suggested by \cite{Thompson2020} to result in better measurements for individual cable modems leads to better data, but considerable market invest and time are required for such a change.
Perhaps as an intermediate step upgrading the monitoring tool could be a more viable option: Other ISPs\footnote{\url{https://medium.com/tele2techblog/great-insights-in-hfc-networks-from-pre-equalisation-data-6b8cab2c1dab 
}} use monitoring tools that intrinsically collect more data, which could further enhance the results if available.

Like in regular ethernet networks, \cite{Dusia2016} HFC could  be inspired by some of the recent advances in fault localization methodologies in internet-protocol (IP) based networks.


\section{Conclusion}\label{sec13}
Enhancing the fault-finding process with machine-learning enhanced models can improve the time to resolution as technicians do not need to follow a lengthy fault-finding process: The best model LightGBM, improves precision@k more than 2.3 times over the baseline for a k of 1.

Furthermore, we have shown that predicting faults at future time steps of the network can be helpful to prevent failures in the network before they show customer impact.


\backmatter



\bmhead{Acknowledgments}

The authors would like to thank the ISP for sharing the anonymized data and the operations team for sharing their multi-decades of HFC experience with us.
In particular, we want to thank Michael Gruber for the pictures of corroded connectors.

\section*{Declarations}
The authors do not have any competing interests.
%
%




\bibliography{bibliography.bib}


\end{document}